\documentclass[reprint,amsmath,amssymb,aps,twocolumn,floatfix]{revtex4-2}
\usepackage{graphicx}% Include figure files
\usepackage{dcolumn}% 
\usepackage{bm}% bold math
\usepackage{relsize}
\usepackage{mathrsfs}
\usepackage{multirow}
\usepackage{textcomp}
\usepackage{textcase}
\usepackage{siunitx}
\usepackage{amsmath}
\usepackage{hyperref}
\usepackage{amssymb}
\usepackage{footnote}
\usepackage{floatrow}
\usepackage{caption, threeparttable}
\usepackage{threeparttable}
\usepackage[utf8]{inputenc}
\usepackage[dvipsnames]{xcolor}
\usepackage[utf8]{inputenc}
\DeclareUnicodeCharacter{2212}{-}

\newcommand{\betac}{\beta_{\text{c}}}

\newcommand{\old}[1]{}

\begin{document}

\preprint{APS/123-QED}

\title{Magnetised relativistic accretion disc around a spinning, electrically
charged, accelerating black hole: case of C-metric}
%Magnetised relativistic accretion disc around a spinning charged accelerating black hole}

\author{Shokoufe Faraji}
 \email{shokoufe.faraji@zarm.uni-bremen.de}
  \affiliation{%
 Center of Applied Space Technology and Microgravity (ZARM), 28359 Germany
}%

 \author{Vladimir Karas}
 \email{vladimir.karas@asu.cas.cz}
  \affiliation{%
Astronomical Institute, Czech Academy of Sciences, Boční II 1401, 141 00 Prague, Czech Republic
} %

\author{Audrey Trova}
 \email{audrey.trova@zarm.uni-bremen.de}
\affiliation{%
 Center of Applied Space Technology and Microgravity (ZARM), 28359 Germany
}%
                                     
%%%%%%%%%%%%%%%%%%%%%%%%%%%%%%%%%%%%%%

\begin{abstract}
This paper examines the general relativistic model of a geometrically thick configuration of an accretion disc around an electrically charged black hole in an accelerated motion, as described by the C-metric family. We aim to study effects of the spacetime background on the magnetised version of the thick disc model via the sequences of figures of equilibrium. While maintaining the assumption of non-selfgravitating (test) fluid, we newly explore the influence of the strength of the large-scale magnetic field with field lines organised over the length-scale of the black hole horizon. We systematically analyze the dependence on a very broad parameter space of the adopted scenario. We demonstrate that the C-metric can, in principle, be distinguished from Kerr black hole metric by resolving specific (albeit rather fine)
features of the torus, such as the location of its center, inner and outer rims, and the overall shape. The analytical set-up can serve as a test bed for numerical simulations.

% My abstract:
%This paper generalises the relativistic accretion thick disc model to the spinning charged accelerating black hole described by the C-metric family. This work aims to study the effects of this background on the thick magnetised version of this model via studying the properties of these equilibrium sequences of magnetised, non-self-gravitating discs. In particular, we analyse the influence of the strength of the magnetic field in this space-time. We show the properties of this relativistic accretion disc model and its dependence on the initial parameters. In principle, this space-time could be distinguished from a Kerr space-time by observing the features of the accretion disc in its vicinity. Besides, this theoretical model can serve as the initial data in numerical simulations.
\end{abstract}

\maketitle

\section{Introduction}
Black holes are the most extreme astrophysical sources of the gravitational field in our Universe: they create an event horizon, hide a singularity, accrete and eject matter from their vicinity, exhibit frame-dragging effects that act on the surrounding particles and fields, produce gravitational waves by violent collisions, {\em etc.}\ \citep[e.g.,][]{1973bhld.book.....D,1989thyg.book.....H,1998LNP...514....3L,2014LNP...876.....R}. An outstanding series of available observations lead to a general agreement that the properties of many astrophysical objects could be best explained in the framework of black-hole accretion disc scenario
\cite{2002apa..book.....F,2008bhad.book.....K}.

One of the theoretical models of accretion discs is the picture of a geometrically thick fluid configuration tori in a stationary, axially symmetrical structure and no magnetic field. This scheme was first introduced in seminal works of early 1970s \citep{1974AcA....24...45A,1978A&A....63..209K,1980AcA....30....1J,1980A&A....88...23P,1980ApJ...242..772A,1981Natur.294..235A,1982MitAG..57...27P,1982ApJ...253..897P}. This model provides a general method to build figures of
equilibrium of the perfect fluid orbiting around an isolated, stationary symmetric black hole. Self-gravity of the fluid was neglected. Later, by confirming the decisive role of magnetic fields for the fluid effective viscosity \cite{1991ApJ...376..214B}, Komissarov proposed an analytical version of the magnetized black-hole torus  \cite{Komissarov_2006}.  The latter work imposes various asumptions, such as the constant specific angular momentum distribution and a strictly toroidal magnetic field configuration, nonetheless, it can serve as a very useful set-up for numerical tests of numerical MHD
schemes \cite{2017MNRAS.467.1838F}.

%%%%%%%%%%%%%%%%%%%%

The above-mentioned approaches consider accretion onto a non-rotating
(Schwarzschild) or rotating (Kerr) black hole in the center of the
torus. In the present work, we focus on the thick disc model in the
space-time of a spinning charged accelerating black hole described by a
generalized family of C-metric \citep{1983GReGr..15..535B,2009esef.book.....S}. The C-metric describes the spacetime of two black holes of equal mass and opposite electric charge. The two black holes undergo acceleration that is directed away at a constant rate. The origin of acceleration can be interpreted as due to a cosmic string that causes a conical singularity and pulls the black holes away from each other. Despite that the presence of singularity complicates the global interpretation, the C-metric can be employed as a toy model to study the effects of an accelerated black hole onto a surrounding test matter within a limited region. Indeed, the origin of this acceleration which is the conical singularity can be replaced for example by a magnetic flux tube \cite{1994PhRvD..49.2909D}.

Originally, the C-metric belongs to a large class of exact solutions discovered by Levi-Civita  \cite{1918NCim...16..105L}. By means of a series of different, rather cumbersome transformations of the Pleba\'nski-Demia\'nski class
of electrovaccuum spacetimes \citep{Plebanski1976}, one could obtain the spinning, charged C-metric. Furthermore, Hong and Teo expressed the metric in a factorized version that is easier to work and can be presented in
Boyer-Lindquist-type coordinates \citep{2003CQGra..20.3269H,2005CQGra..22..109H}. In fact, the maximal analytical extension of the line element describes two causally disconnected black holes accelerating in the opposite directions \citep{PhysRevD.2.1359}.

%%%%%%%%%%%%%%%%%%%%%%%%%%%
In this paper we investigate this background with the aim of study the properties of magnetised tori and the morphology of the equipotential surfaces.

In this paper we investigate the background space-time of C-metric with the aim of studying the properties of magnetised tori and the morphology of the equipotential surfaces. There are different motivations to consider this research. First of all, this is an exact solution to Einstein's field equation, which is worth understanding for its own sake. Besides, most of the works in the astrophysics area have been done by assuming the Schwarzschild or Kerr metrics are the best description
of astrophysical compact objects in the relativistic astrophysics area.

Secondly, astrophysical observations may not all fit within the general theory of relativity by employing just to a restricted family of Kerr spacetime \citep{PhysRevD.78.024040,2019MNRAS.482...52S, 2002A&A...396L..31A}. It appears entirely natural to explore departures due to small electric charge and translatory motion of an accelerated black hole.

C-metric allows for electric charge and acceleration parameters. In fact, even if astrophysical black holes are assumed to be neutralized by their environment, a tiny net equilibrium charge may remain  \cite{PhysRevD.10.1680,2018MNRAS.480.4408Z}. Besides, considering the acceleration parameter in this setup may provide first steps to a (semi-)analytical description of stellar-mass black holes that have received recoil velocity at their formations. In fact, there is a widespread agreement that the birth kicks of black holes are necessary to explain the large distances above the Galactic plane achieved by some binaries \cite{2012MNRAS.425.2799R} and caused the black hole to accelerate within a local cosmological medium.

In this perspective, the family of C-metric could be a hypothetical candidate for objects that exist in nature. In order to investigate this question, the study of its fingerprint in the observational data can be a proper first step along with analytical analysis. On the other hand, it seems the only source of information that we have in the strong gravity regime is coming from its environment, like from the shadow or accretion discs, especially with the advent of horizon-scale observations of astrophysical black holes. There is a vast literature on the lensing in this metric \citep[e.g.][]{doi:10.1142/S0218271815420249,GIBBONS2016169,Alawadi_2020,Frost:2020zcy}, and its radiative nature \cite{1999PhRvD..60d4004B,2003PhRvD..68l4004P}. To our knowledge, the influence of generalized C-metric parameters on accretion discs has not yet been properly examined.

%,Frost:2020zcy, doi:10.1142/S0218271815420249
%\citep{GIBBONS2016169,Alawadi_2020}

% \citep{doi:10.1142/S0218271815420249,GIBBONS2016169,Alawadi_2020,Frost:2020zcy}

%cite{2012CQGra..29c5010K} for acceleration

%%%%%%%%%%%%%%%%%%%%%%%%%%%

The organization of the paper is as follows: the space-time is briefly explained in \ref{sec2}. The relativistic magnetised tori presented in section \ref{sec3}. The results and discussion are presented in Section \ref{sec4}, and the conclusions are summarised in Section \ref{sec5}.
In this paper, the geometrized units where $c=1$, and $G=1$, also the signature $(-+++)$ are used.
%%%%%%%%%%%%%%%%%%%%%%%%%%%%%%%%%%%

%Recently Alrais Alawadi et al. \cite{2020arXiv200603376A} rederived, in Boyer-Lindquist-like coordinates, the coordinate radius of the photon sphere and they also determined the coordinate angle of what we will refer to as the photon cone. In addition they performed a stability analysis and found that the circular null geodesic at the intersection of the photon sphere and the photon cone is unstable with respect to radial perturbations. Actually, all lightlike geodesics in the photon sphere are unstable with respect to radial perturbations \cite{2020arXiv201011908F}.

%%%%%%%%%%%%%%%%%%%%%%%%%%%%%%%%%%%%%%%%%%%
%%%%%%%%%%%%%% SEC 2

\section{Spinning charged C-Metric}\label{sec2}
The family of C-metric has accelerating nature and is considered as describing accelerating black holes \citep{PhysRevD.2.1359}. The spinning charged C-metric in Boyer-Lindquist-type coordinates \citep{2005CQGra..22..109H} reads as  

\begin{align}\label{eq:metric}
&ds^2=\frac{1}{\Omega^2}\left(-\frac{f}{\Sigma}\left[dt-a\sin^2\theta\frac{d\varphi}{K}\right]^2\right.\nonumber\\
&\left.+\frac{\Sigma}{f}dr^2+\Sigma r^2\frac{d\theta^2}{g}+\frac{g\sin^2\theta}{\Sigma r^2}\left[adt-(r^2+a^2)\frac{d\varphi}{K}\right]^2\right),\  
\end{align} 
with the metric functions

\begin{align}
    &\Omega=1+\alpha r \cos\theta,\\
    %&H=1+\alpha r \cos\theta,\\
    &f(r)=\left(1-\alpha^2r^2\right)\left(1-\frac{2m}{r}+\frac{e^2+a^2}{r^2}\right),\\
    &g(\theta)=(e^2+a^2)\alpha^2 \cos^2\theta+2m\alpha \cos\theta+1,\\
    %&h(\theta)=1+2 \alpha m \cos\theta+\left[\alpha^2(e^2+a^2)\right] \cos^2\theta,\\
    &\Sigma(r,\theta) = \frac{a^2}{r^2}\cos^2\theta+1,\\
    &\xi=\alpha^2(e^2+a^2)+1,\\
    &K=\xi+2m\alpha,
\end{align}
where $t\in (-\infty,+\infty)$, $\theta\in (0, \pi)$, $r\in (0, +\infty)$. The metric has four independent parameters: the mass $m$, the electric charge $e$, the rotation $a$, and the so-called acceleration parameter $\alpha$.

In this metric, $r=0$ is the curvature singularity, and there is also a conical singularity on the $\theta$ axis. In fact, the conical deficit is associated with the presence of a cosmic string. Since the deficit along both axis $\theta=0$ and $\theta=2\pi$ are not the same, this imbalance tension is the origin of the driven acceleration. However, the parameter $K$ in the metric regulates the distribution of tensions along either axis and allows $\varphi$ to be $2\pi$-periodic. It is also worth mentioning that a negative deficit is also possible; however, theoretically this would be sourced by a negative energy object.

\begin{figure}
   \centering
   \includegraphics[width=7cm]{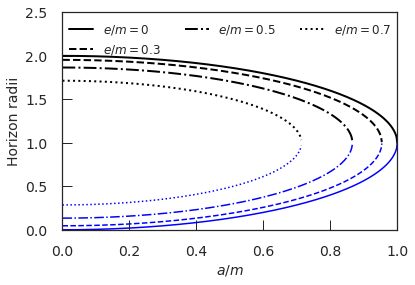}
    \caption{Place of horizons as a function of rotation parameter $a$, for different charge parameter $e$.}
    \label{fig:h1}
\end{figure}

Almost all analysis considering the family of C-metric are revolved around the coordinate ranges, which are dictated by the metric functions and their root configurations. First of all, the conformal factor $\Omega$ determines the location of the boundary

\begin{align}
  r_b = -\frac{1}{\alpha\cos \theta}.
\end{align}
In addition, the roots of metric function $f(r)$ correspond to horizons. Thus, $f(r)$ should have at least one root for $r\in(0,\frac{1}{\alpha})$ to have a black hole in the space-time. In general, mostly with non vanishing charge $e$, generic configurations have different distinct horizons. In fact, they happen to be in a pair where they are known as inner and outer horizons. Like the regular Reissner-Nordstroem solution, they typically approach one another and vanish for relatively high charge. Furthermore, when the acceleration horizon is present, there is a second outer acceleration horizon, and both intersect with the boundary. In general, the pairs of horizons separated the space-time into different regions which share the same signature. For astrophysics point of view, we are interested in studying the accretion disc in the outer communication region between the outer horizon and the acceleration horizon. In Figure \ref{fig:h1} the place of inner and outer horizon have presented for the chosen parameters. As we see, by increasing $e$ the place of horizons become closer to each other, and to the black hole. However, since the accelerating horizon depends on $\alpha$, the valid region becomes wider. The same behaviour expected for increasing $a$, as the metric function $f(r)$ is symmetric in parameters $e$ and $a$.

\begin{figure}
    \centering
        \includegraphics[width=6cm]{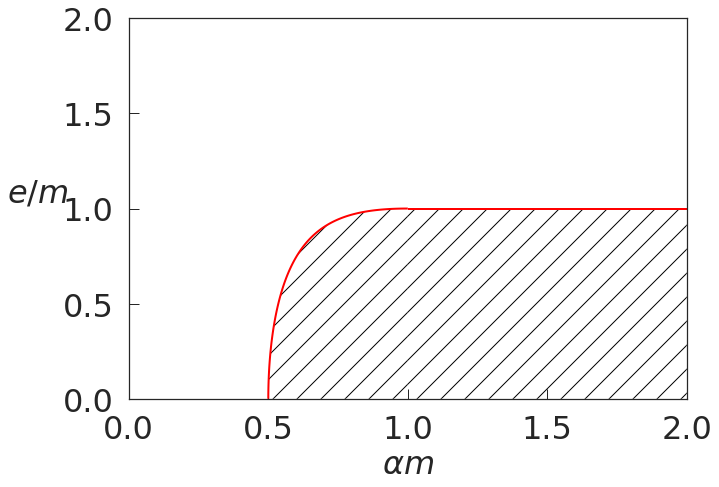}

        \includegraphics[width=6cm]{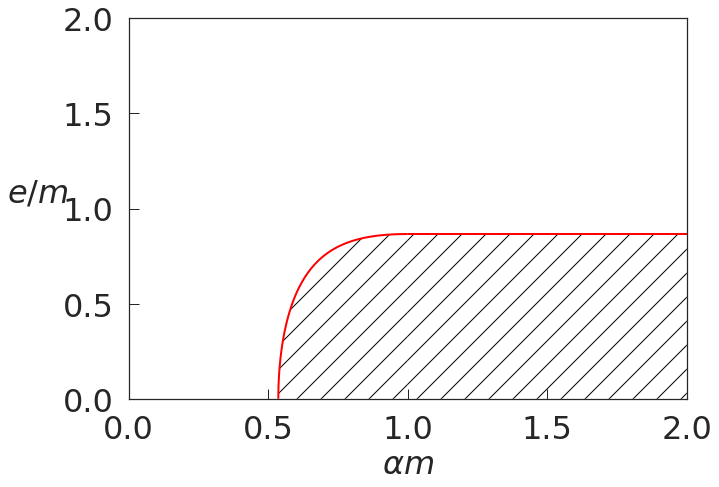}
           \caption{Allowed parametric regions for the spinning charged C-metric as a function of $e$. The regions marked out with hatching corresponds to the forbidden regions. First one represent the result for $a=0$, and the second one for $a=0.5$.}
    \label{fig:regione}
\end{figure}

\begin{figure}
    \centering
    \includegraphics[width=6cm]{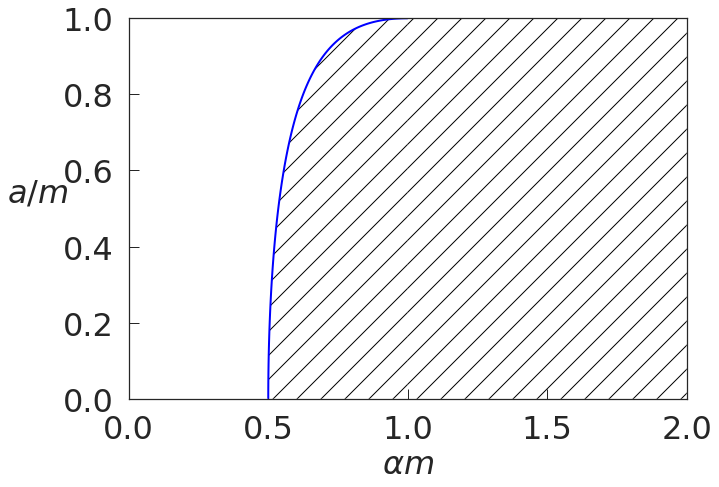}
    \includegraphics[width=6cm]{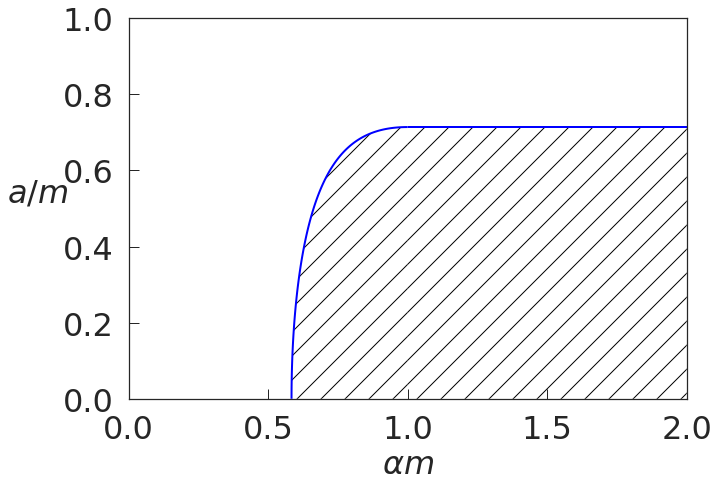}
    \caption{Allowed parametric regions for the spinning charged C-metric as a function of $a$. The regions marked out with hatching corresponds to the forbidden regions. First one represent the result for $e=0$, and the second one for $e=0.7$. }
    \label{fig:regiona}
\end{figure}

\begin{figure*}
   \centering
   \begin{tabular}{cc}
\includegraphics[width=6cm]{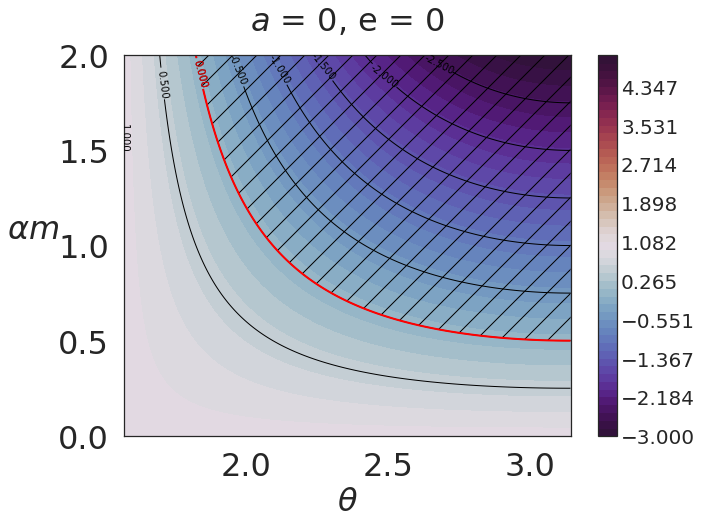}&
\includegraphics[width=6cm]{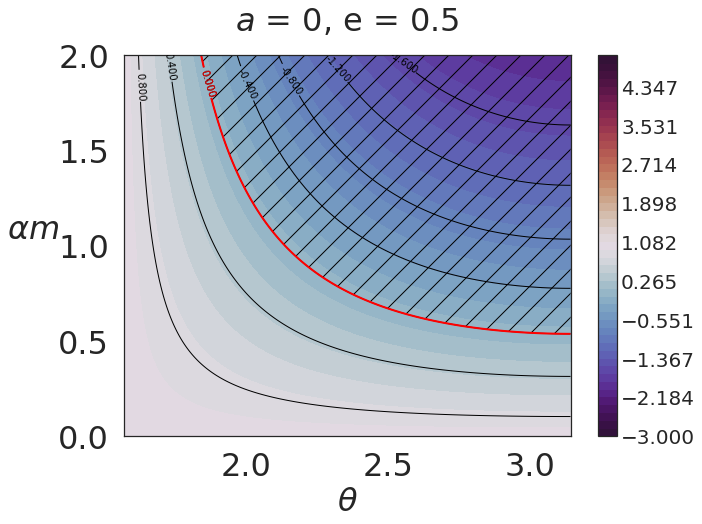}\\
\includegraphics[width=6cm]{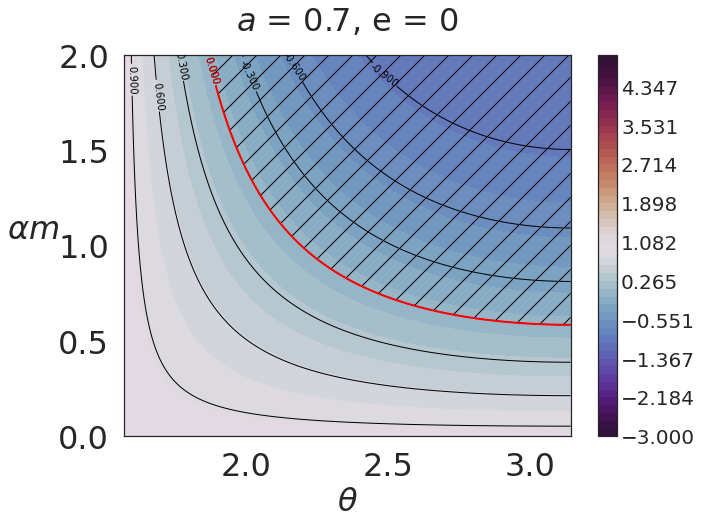}&
\includegraphics[width=6cm]{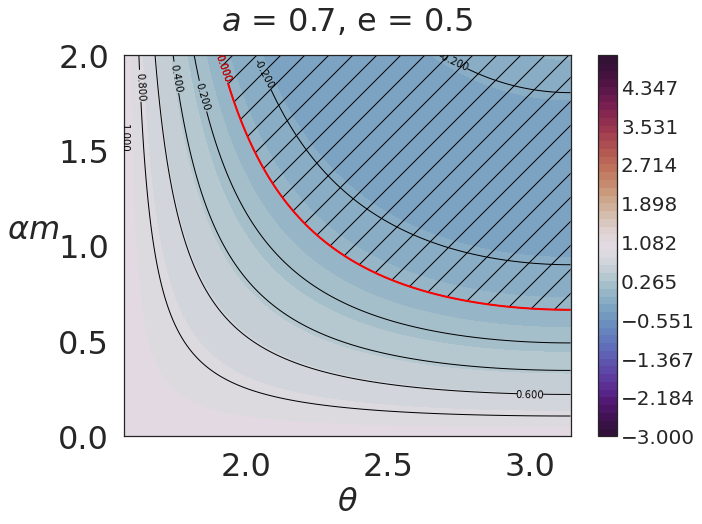}\\
\end{tabular}
\caption{Plots of metric function $g(\theta)$. The red line corresponds to $g(\theta)=0$, and the hatched region shows $g(\theta)<0$.}
    \label{fig:gplot}
\end{figure*}

Finally, from the analyzing the $\theta$-coordinate, the metric function $g(\theta)$ should have positive roots in $[0,\pi]$. Therefore, it requires to have

\begin{equation}
  e^2+a^2\leq m^2,  
\end{equation}
and the following condition should be fulfilled

\begin{align}\label{regioneq}
 2m\alpha\leq
   \left\{
  \begin{array}{@{}ll@{}}
  2\sqrt{\xi-1} & \xi>2, \\
    \xi & 0<\xi\leq 2.
    \end{array}\right.
\end{align}
Figures \ref{fig:regione} and \ref{fig:regiona} show different parametric setups. The hatched parts are the excluded regions by the equation \eqref{regioneq} for chosen parameters. We see that this condition acts as an upper bound on the rotation or the acceleration for relatively small parameters. Figure \ref{fig:gplot} shows the metric function $g(\theta)$ for chosen parameters. As it has shown, the forbidden region is the hatched part, which shrinks as $a$ and $e$ increase since $g(\theta)$ is also symmetric with respect to $a$ and $e$.

\begin{figure*}
\centering
\begin{tabular}{ccc}
 \includegraphics[width=6cm]{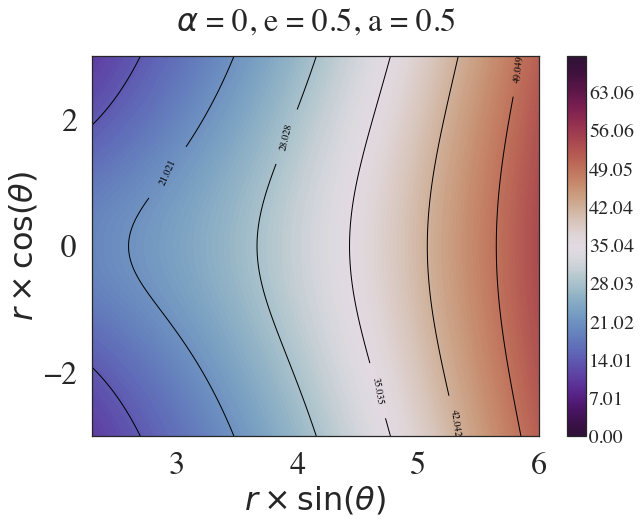}&
\includegraphics[width=6cm]{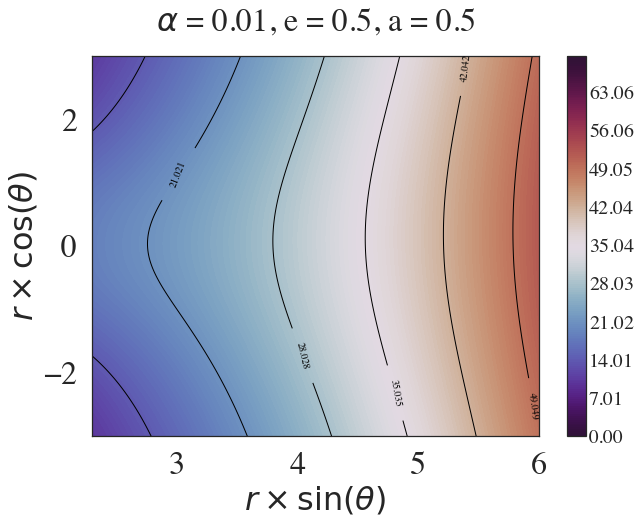}&
\includegraphics[width=6cm]{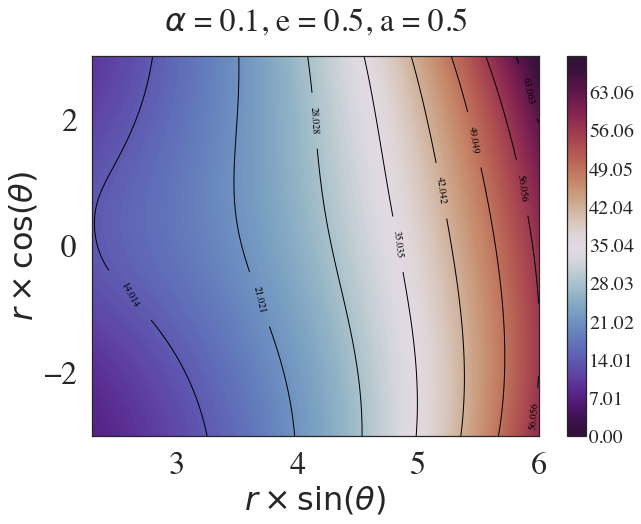}\\
\includegraphics[width=6cm]{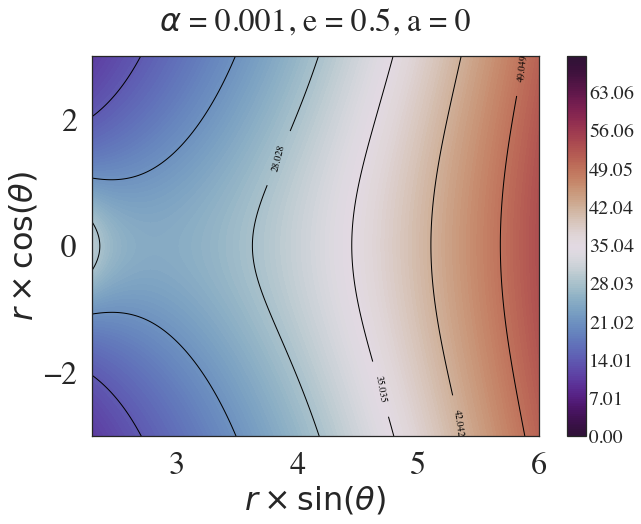}&
\includegraphics[width=6cm]{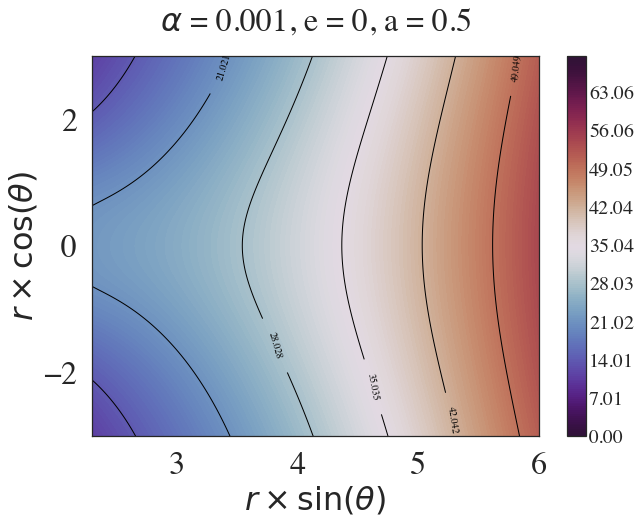}&
\includegraphics[width=6cm]{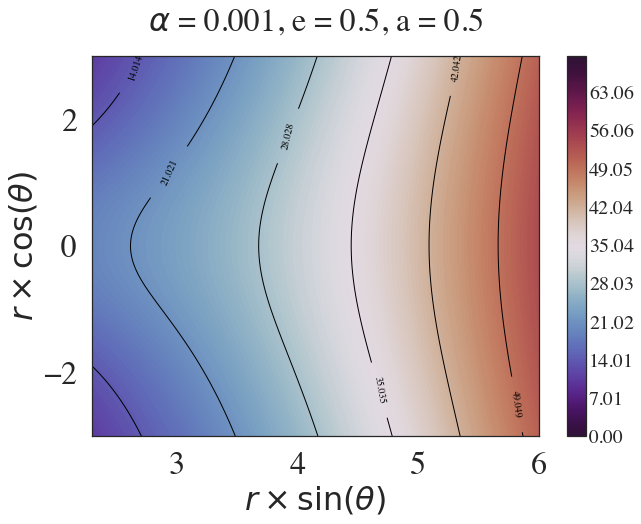}
\end{tabular}
\caption{Von Zeipel cylinders with respect to the stationary observers. The plots show poloidal sections across the constant R surfaces, where the circular time-like motion of the fluid is possible. Colours correspond to different values of the radius, as listed on the colour-bar to the right of each panel. Selected contours are indicated with black lines.}
    \label{cyl}
\end{figure*}

Before we describe the construction of the relativistic thick disc model, we briefly discuss the modified von Zeipel radius (or radius of gyration) as an important concept of the thick disc model \citep{1924MNRAS..84..665V}. This radius determines surfaces of constant $\mathcal{R}$, which for an axisymmetric and the stationary metric is defined as

\begin{equation}\label{vonc}
\mathcal{R}=\frac{g^2_{\varphi \varphi}}{g^2_{t\varphi}-g_{tt}g_{\varphi \varphi}},
\end{equation}
with respect to the stationary observers, and known as von Zeipel cylinders \citep{PhysRevD.47.1440,1991MNRAS.250....7C}. In Figure \ref{cyl}, we see the effect of parameters on this radius for different sets of $\alpha$, $e$ and $a$ in this space-time. In brief, this radius helps to analyze circular particle motion and provides an intuitive image of them in this space-time; besides, this radius is related to the inertial forces. In this concept using von Zeipel theorem, we can conclude that for a constant angular momentum distribution, the surface of constant $\mathcal{R}$ and constant $\Omega$ coincide. This model is summarized in the next section.

%\begin{figure*}
 %  \centering
  % \begin{tabular}{cc}
%\includegraphics[width=8cm]{RadiusVZalpha=0.001e=0a=0.png}&
%\includegraphics[width=8cm]{RadiusVZalpha=0.005e=0a=0.png}\\
%\end{tabular}
%\caption{Zeipel radious. for different \alpha and e=a=0}
 %   \label{cyl}
%\end{figure*}

%%%%%%%%%%%%%%%%%%%%%%%%%%%% %%%%%%%%%%%%%%%SEC 3

\section{Relativistic Thick disc model}
\label{sec3}
The relativistic thick disc model considers perfect fluid equilibria of matter with the barotropic equation of state orbiting in the azimuthal direction. This equlibrium tori can study objects with sufficient radial pressure gradients which tends to enlarge the vertical size of the disk. Additionally, it models the radiatively inefficient and non self-gravitating accretion discs without accretion flow analytically based on Boyer's condition \footnote{Boyer's condition states the boundary of any stationary and barotropic perfect fluid body is an equipotential surface.}. Within the assumption of this model, the four-velocity and stress-energy tensor are determined by
%t assumption is not fulfilled close to the cusp
\begin{align}
    &u^{\mu}=(u^t,0,0,u^{\varphi}),\\
    &T^{\mu}{}_{\nu}=wu_{\nu}u^{\mu}+\delta^{\mu}{}_{\nu}p,\
\end{align}
where the dissipation processes are neglected \citep{1989rfmw.book.....A}. Here $w$ is the enthalpy, and $p$ is the pressure. The relativistic Euler equation by considering the projection of conservation of stress-energy tensor into the plane normal to four-velocity is written as \cite{1978A&A....63..221A},
%\begin{align}\label{7}
 %   \frac{1}{w}\nabla{}_{i}p=-\nabla_{i}\ln{u_t}+\frac{\Omega}{1-\Omega\ell}\nabla{}_{i}\ell, 
%\end{align}
%where $i$ is $r$ or $\theta$. For a barytropic equation of state, $p=p(\rho)$, where $\rho$ is the internal energy density, this leads to 

\begin{align}\label{maintori}
    \int_{p_{\rm in}}^{p}{\frac{{\rm d}p}{w}}=-\ln{|u_t|}+\ln{|(u_t)_{\rm in}|}+\int_{\ell_{\rm in}}^{\ell}{\frac{\Omega{\rm d}\ell}{1-\Omega\ell}},
\end{align}
where $\ell=-\frac{u_{\varphi}}{u_{t}}$, and $\Omega=\frac{u^{\varphi}}{u^{t}}$, also the subscript \textit{in} refers to the inner edge of the disc. The integrability condition exhibits $\Omega=\Omega(\ell)$, and this relation fulfills the relativistic von Zeipel theorem for a toroidal magnetic field \cite{2015GReGr..47...44Z}. This theorem states that the surfaces of constant pressure coincide with the surfaces of the constant enthalpy if and only if surfaces of constant $\Omega$ and constant $\ell$ coincide. In another word; the surfaces "$\mathcal{R} = const.$" coincide with the surfaces "$\Omega= const.$", as mentioned earlier. %Nevertheless, vanishing magnetic field, the whole surfaces of equal $\Omega$, $\ell$, and $w$ coincide.

In what follows, we continue with the generalization of the thick disc model by considering the magnetic field that has been developed by \cite{Komissarov_2006}. %\footnote{which states the toroidal magnetic field, the surfaces of constant $p$ coincide with constant $w$ if and only if constant $\Omega$ and constant $\ell$ coincide \cite{1924MNRAS..84..665V, 2015GReGr..47...44Z}. However, for the non magnetised version, the surface of equal $\Omega$, $\ell$, $p$ and $w$ all coincide \cite{1978A&A....63..221A}.}.
%%%%%%%%%%%%%%%%%%%

\subsection{\label{sec3.1}Magnetised version}

 % (for this reason they are also referred to as accreting tori ). These

The dynamical evolution of the disk model in the presence of magnetic field is governed by the following conservation laws \citep{1989rfmw.book.....A,1978srfm.book.....D},

\begin{align}\label{eq:Property}
&\nabla_{\nu}T^{\nu\mu}=0 \,,\\
&\nabla_{\nu}\left(\rho u^{\nu}\right)=0\,,\\
&\nabla_{\nu}(b^{\nu} u^{\mu}-b^{\mu} u^{\nu})=0\,,
\end{align}
The energy-momentum conservation, baryon conservation, and induction equation, respectively. Here, $\rho$ is the mass density, $b^{\mu}$ are the components of magnetic field and is given by magnetic pressure in the fluid frame as $|b|^2=2p_{\rm m}$ \citep{1989rfmw.book.....A,1978srfm.book.....D}. The total energy-momentum tensor is given by

\begin{align}
T^{\nu \mu}=\left(w+|b|^{2}\right) u^{\nu} u^{\mu}+\left(p_{\rm gas}+\frac{1}{2} |b|^{2}\right) g^{\nu \mu}-b^{\nu} b^{\mu},
\end{align}
where $p_{\rm gas}$ is the gas pressure, and as mentioned the dissipation processes are negligible.
%\textbf{We need to mention that the fluid enthalpy gives by $w=\rho h$, where $h$ is the specific enthalpy. As we assume following the preceding works the fluid is thermodynamically nonrelativistic, we should choose $h=1$. This is supported by the fact that we are using a polytropic equation of state of the same form to link the pressure to either the enthalpy or the rest-mass density. It is worth noticing that for a circular rotating perfect fluid, the shapes and location of the equipressure surface $p(r, \theta)$ = const. are characterized by the assumed angular momentum distribution independently of the equation of state and the assumed entropy distribution \cite{1980AcA....30....1J}.} 
By assuming the model is axisymmetric and stationary

\begin{align}
u^r=u^{\theta}=b^r=b^{\theta}=0.
\end{align}
The equation \eqref{maintori} is generalized to
\begin{align}\label{9}
    \int_0^{p}{\frac{{\rm d}p}{w}}+\int_0^{\Tilde{p}_{\rm m}}{\frac{{\rm d}\Tilde{p}_{\rm m}}{\Tilde{w}}}&=-\ln{|u_t|}-\ln{|(u_t)_{in}|}+\int_{\ell_{in}}^{\ell}{\frac{\Omega{\rm d}\ell}{1-\Omega\ell}}.
\end{align}
where $\Tilde{p}_{\rm m}=\mathcal{L}p_{\rm m}$, and $\Tilde{w}=\mathcal{L}w$, with $\mathcal{L}=g^{2}_{t \varphi}-g_{tt} g_{\varphi \varphi}$. To solve this equation, we assume the following relations \citep{Komissarov_2006}

\begin{align}
p=K w^{\kappa}, \quad \tilde{p}_{\mathrm{m}}=K_{\mathrm{m}} \tilde{w}^{\eta}
\label{eq:polytrops}
\end{align}
where $K$, $K_{\mathrm{m}}$, $\kappa$ and $\eta$ are constants. Further, $\tilde{p}_{\mathrm{m}}$ can be rewritten in terms of the magnetic pressure as $p_m=K_m {\cal{L}}^{\eta-1}w^{\eta}$. The equation \eqref{9} allows that on the surface of the disc and its inner edge the pressures vanish by choosing the integration constant. Thus, the equation \eqref{9} is integrable

\begin{align}\label{eq:FinalEq2}
W-W_{\mathrm{in}}+\frac{\kappa}{\kappa-1}\frac{p}{w}+\frac{\eta}{\eta-1} \frac{p_m}{w}=\int_{\ell_{in}}^{\ell}{\frac{\Omega{\rm d}\ell}{1-\Omega\ell}},
\end{align}
where $W=\ln|u_t|$. This equation implies $\Omega=\Omega(\ell)$ \citep{1978A&A....63..221A}; therefore, by choosing $\Omega=\Omega(\ell)$ equipotential surfaces $W$ and pressure $p$ will be determined. In addition, to fix the geometry of the equipotential surfaces, it is required to choose angular momentum distribution $\ell$. In this work, we consider the constant angular momentum $\ell_0$, while choosing constant distribution profile causes the right-hand side of the equation \eqref{eq:FinalEq2} vanish and by specifying $W_{\rm in}$ one can obtain the solutions. Finally, the total potential is given by

\begin{align}\label{K33}
    W(r,\theta)=\frac{1}{2}\ln|\frac{\mathcal{L}}{g_{\varphi\varphi}+2\ell_0g_{t\varphi}+\ell^2_0g_{tt}}|.
\end{align}
However, $\ell_0$ should be chosen in the interval between the marginally stable orbit $l_{ms}$ and the marginally bound orbit $l_{mb}$, to constructing a finite-size disc \cite{1978A&A....63..221A}, hence 

\begin{align}
    \left\{
  \begin{array}{@{}ll@{}}
  W_{\rm in}\leq W_{\rm cusp} & \text{if}\ |\ell_{ms}|<|\ell_0|< |\ell_{mb}|, \\
  W_{\rm in}<0 & \text{if}\ |\ell_0|\geq |\ell_{mb}|.
    \end{array}\right.
\end{align}
The cusp point is defined as the circle in the equatorial plane on which the pressure gradient vanishes and the specific angular momentum of the disc equals the Keplerian angular momentum. Besides, the pressure at the centre of the disk, denoted by index $c$, is determined by

\begin{align}
    p_c= w_c(W_{\rm in}-W_{c})\left(\frac{\kappa}{\kappa-1}+\frac{\eta}{\beta_{c}(\eta-1)}\right)^{-1},
    \label{eq:pc}
\end{align}
center of the disk is the place where we have the maximum pressure. Here $\beta=p/p_{m}$ is the magnetization parameter. The variables of model are then $u^t$, $u^{\varphi}$, $b^t$ and $b_{\varphi}$. In addition, $W$, $w$, $p$, and $p_{\rm m}$ computed utilizing equations \eqref{eq:polytrops}, \eqref{eq:FinalEq2} and \eqref{K33}.

%In conclusion, the equation of state induces $K$ and $K_{\rm m}$ \cite{Komissarov_2006} thenceforth the equations \eqref{eq:FinalEq2} and \eqref{K33} provide the solutions for $W$, $p$ and $p_m$ and other variables are followed.

%{\color{blue} Let us note that, as mentioned in previous section, the disk is not in the equatorial plane, then finding conditions of existence of equipotential surfaces are more challenging as in the Kerr metric. One more extra equations as to be fulfilled to be able to find the right parameters allowing the existence of the disc.}

%\begin{align}
   % \Omega=&-\frac{g_{t\varphi}+g_{tt}\ell_0}{g_{\varphi\varphi}+g_{t\varphi}\ell_0},\\
    %(u^t)^{-1}=&-u_t(1-\ell_0\Omega),\\
    %u^{\varphi}=&\Omega u^t,\\
    %b^{\varphi}=&\pm \sqrt{\frac{2p_{\rm m}}{\mathcal{A}}},\\
    %b^t=&\ell_0b^{\varphi}.\
%\end{align}

%%%%%%%%%%%%%%%%%%%%%%%%%%%%%%%%%%%% LATER
%\subsection{Non-constant angular momentum}\label{4}
%In what follows, we briefly explain two models for angular momentum distribution introduced in \citep{2015MNRAS.447.3593W} and \citep{2009A&A...498..471Q}, which we consider in this paper.
%%%%%%%%%%%%%%%%%%%%%%%%%%%%%%%%%%

\begin{figure*}
  \centering
\begin{tabular}{cc}
\includegraphics[width=6.5cm]{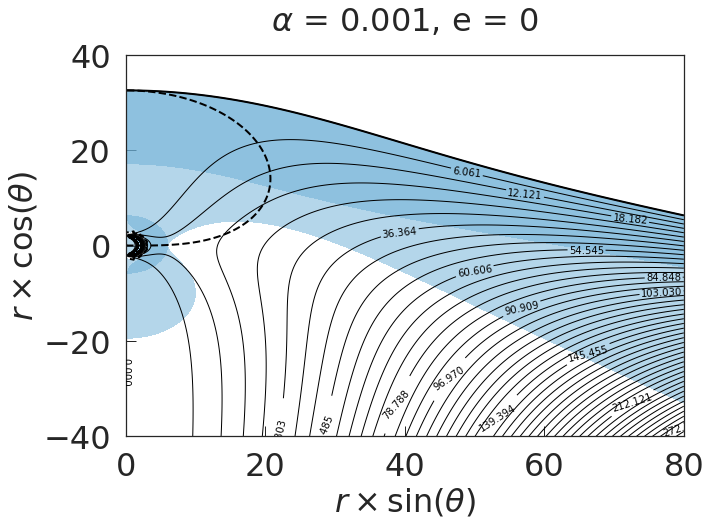}&
\includegraphics[width=6.5cm]{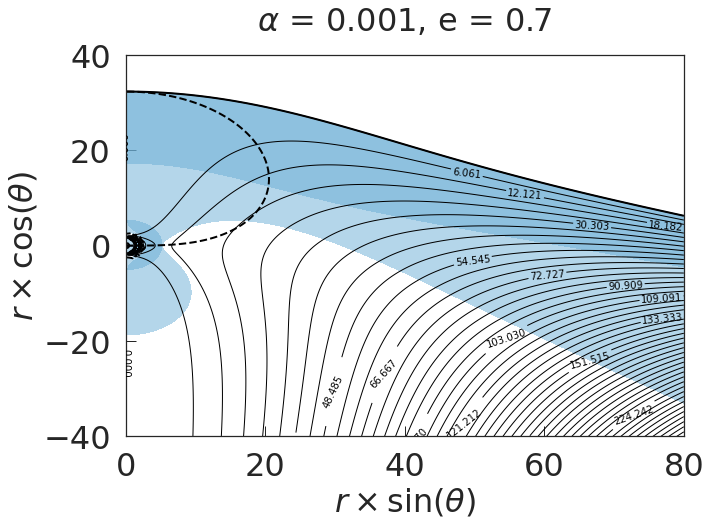}\\
\includegraphics[width=6.5cm]{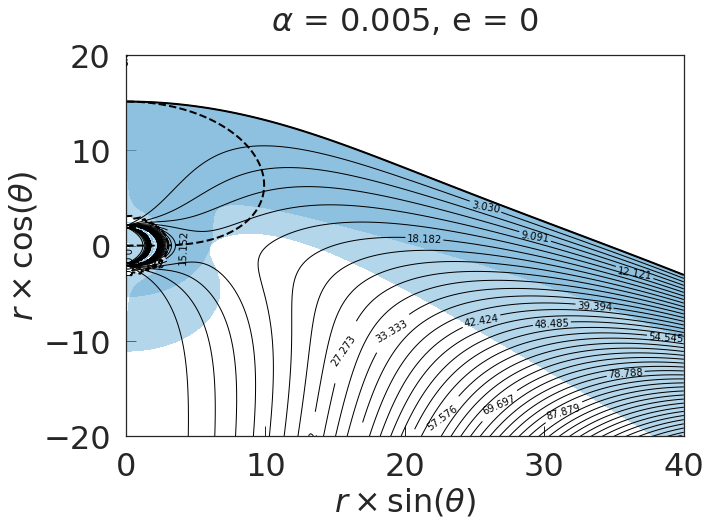}&
\includegraphics[width=6.5cm]{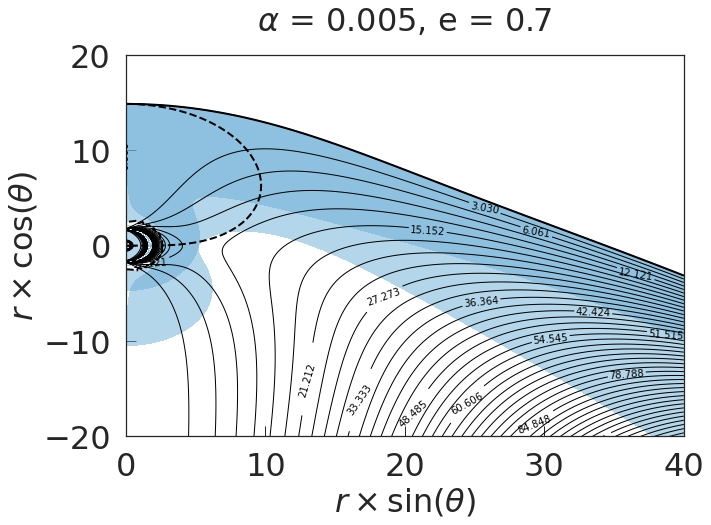}\\
\end{tabular}
    \caption{Possible region for having the thick disc model. The dashed curves shows when $\partial_{\rm r}W=0$ and $\partial_{\rm \theta}W=0$. The dark blue region shows area where the condition for a maximum of $W(r,\theta)$ are fulfilled. The white area depicts where the condition for a minimum of $W(r,\theta)$ are fulfilled.} %On the top left graph: the crossing thick line point a set of coordinates corresponding to a choice of ($r_c,\theta_c$). The red curve shows a specific contour of $L_K$.}
    \label{fig:figsolre} % I can do without the label too
\end{figure*}

In what follows we present the thick disc model in this space-time; however, because of the conical deficit, the disc does not lie on the equatorial plane and finding conditions of existence of equipotential surfaces are rather a challenging task.

%%%%%%%%%%%%%%%%%%%%%%%%%%%%%%%%%%%%%%%%%%%%%%% Sec 4

\section{Results and discussion}\label{sec4}

In this section, we analyzed the impact of the different parameters of the model on the morphology of the equipotential surfaces.

We started by examining the possibility of having solutions for this disc's model rely on the variation of the parameters. To have a better insight on the role of acceleration parameter, first we consider non-spinning case. Figure \ref{fig:figsolre} shows the regions where we may have a disc in the non-spinning charged set-up.

In the panel, the intersection of the dashed curve with the white and the dark-blue areas show the possible places for choosing the centre of the disc, and the cusp point, respectively. As we see the acceleration parameter $\alpha$ plays a crucial role in the existence of solutions. As a result, as  $\alpha$ even slightly increases, almost the possibility of having solutions decreases dramatically. Therefore, we can build this disc's model only for relatively small acceleration. On the other hand, the charge parameter $e$ has an imperceptible impact, and positively contributes to having solutions. Especially, the charge's effect manifests clearly when $\alpha$ is relatively large. For example, with a comparison in the second row for relatively large $\alpha$ and vanishing $e$, we do not have a solution while by increasing $e$, we obtain solutions.

Figure \ref{fig:figsolre} also shows that the distance between the center and the cusp point changes as a monotonic decreasing function of $\alpha$. This leads to the larger disc structure for smaller rotation parameters $a$. In conclusion, as $\alpha$ increases with a moderate rate, the disc structure becomes smaller and finally, it will vanish. Besides, the deeper analysis of the panel manifests the possibility of the existence of two cusps for some choices of parameters.

\begin{figure*}
  \centering
\begin{tabular}{ccc}
\includegraphics[width=5.5cm]{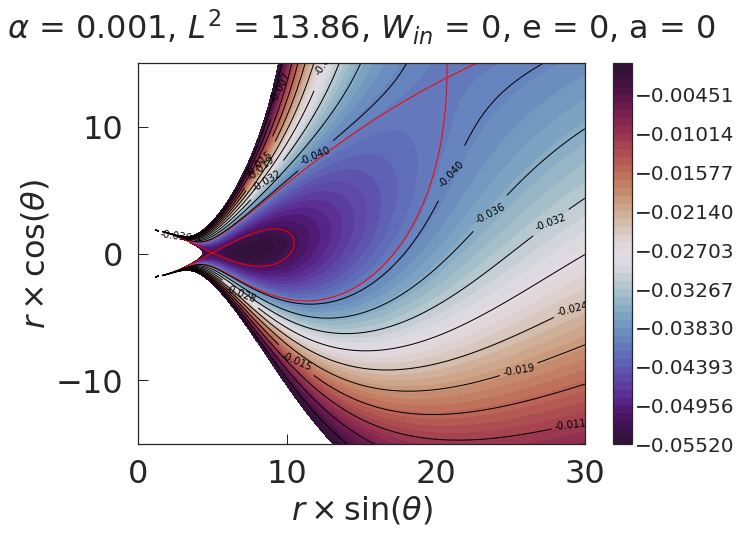}&
\includegraphics[width=5.5cm]{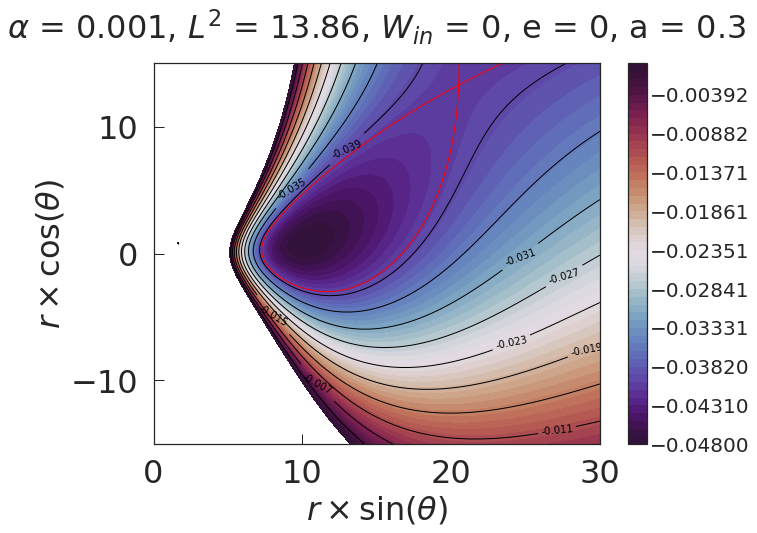}&
\includegraphics[width=5.5cm]{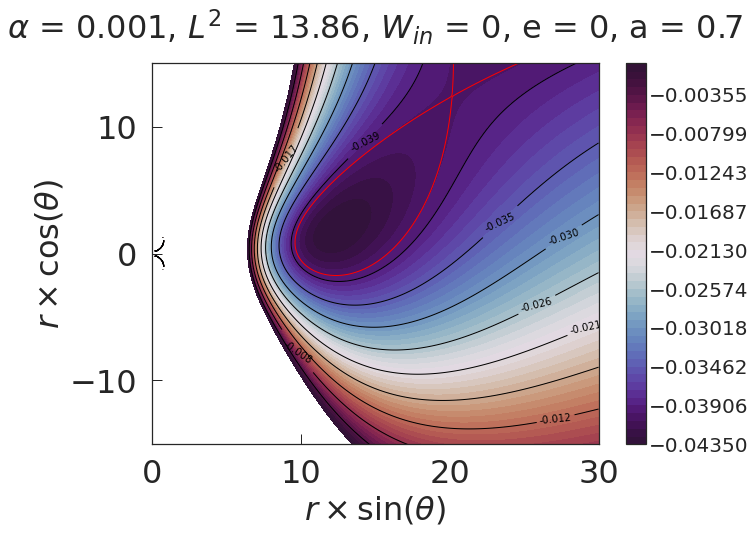}\\
\includegraphics[width=5.5cm]{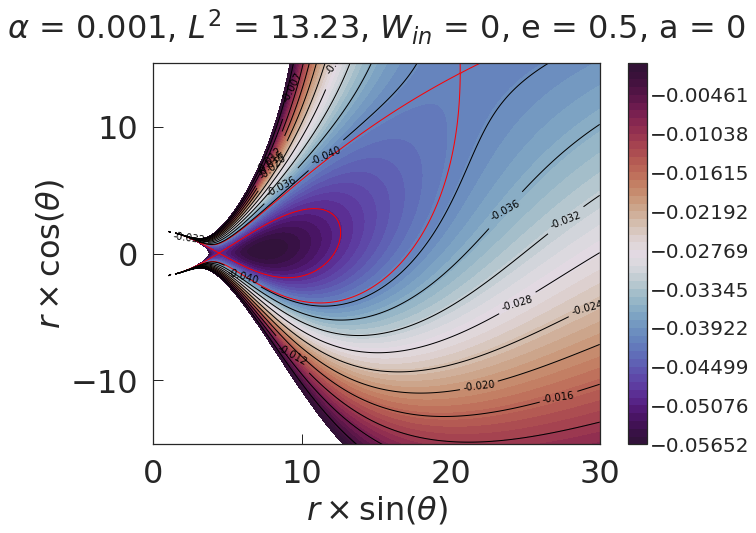}&
\includegraphics[width=5.5cm]{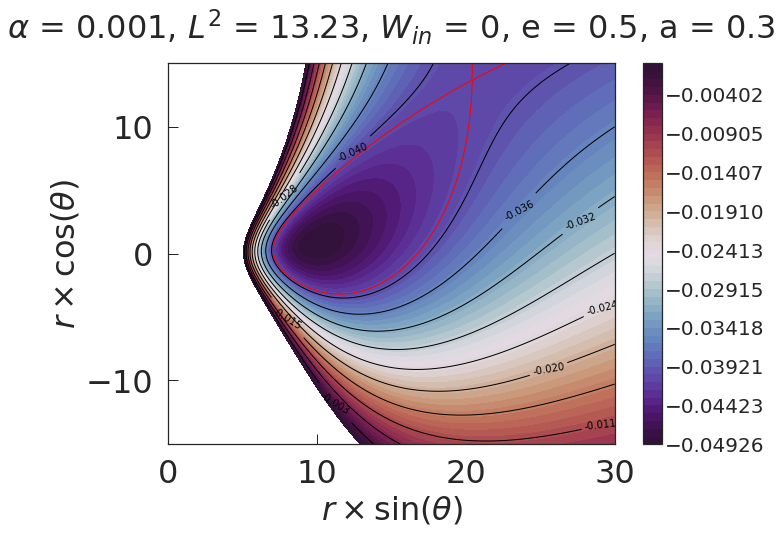}&
\includegraphics[width=5.5cm]{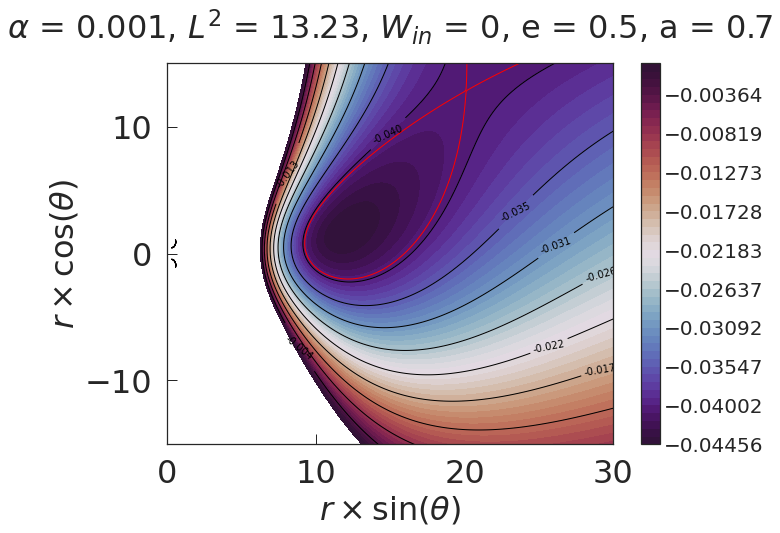}\\
\includegraphics[width=5.5cm]{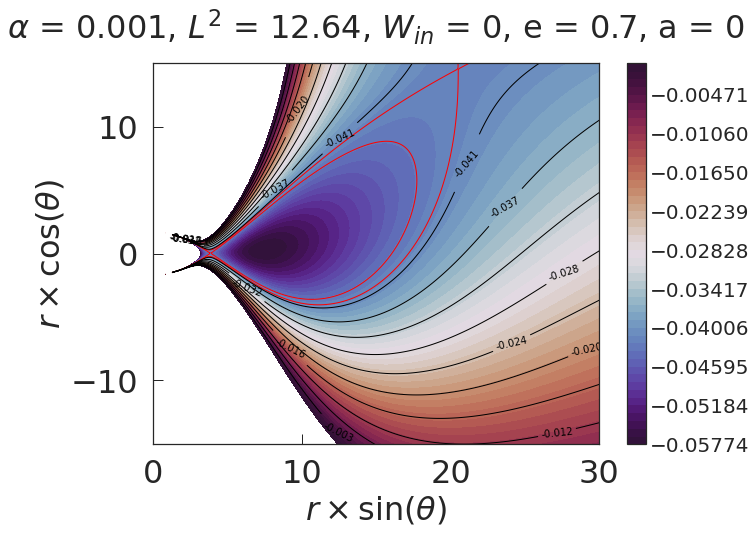}&
\includegraphics[width=5.5cm]{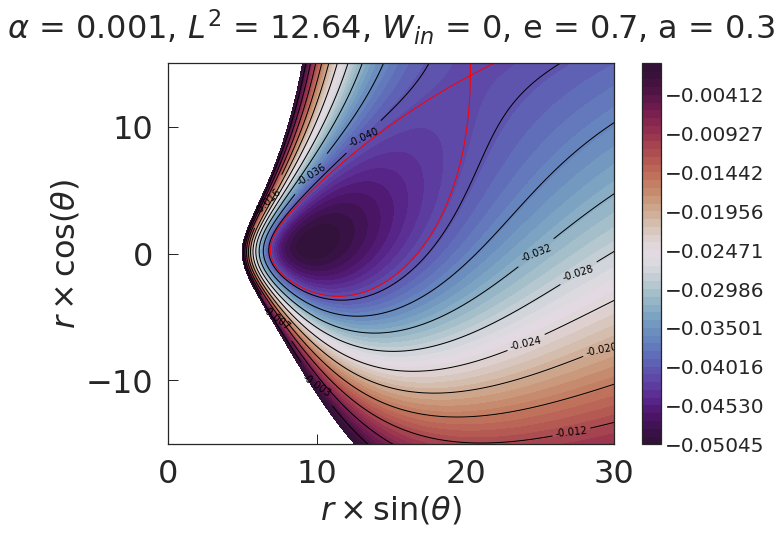}&
\includegraphics[width=5.5cm]{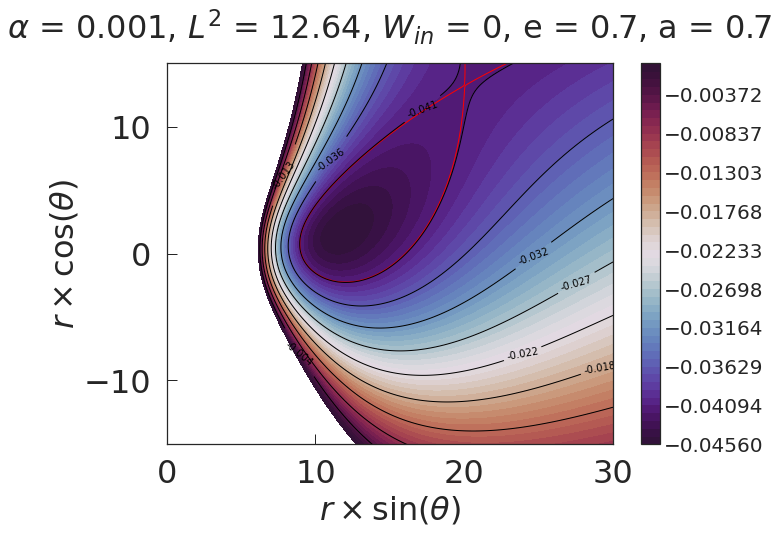}\\
\end{tabular}
    \caption{Contour map of the equipotential surfaces. The red lines show the equipotential corresponding to the inner and outer cusps.}
    \label{fig:figsolre1rot} % I can do without the label too
\end{figure*}

In Figure \ref{fig:figsolre1rot}, by using Figure \ref{fig:figsolre}, we choose a solution possessing an inner cusp, a centre and an outer cusp to construct
the largest possible model. In Figure \ref{fig:figsolre1rot} clearly can be seen for vanishing rotation in the first column, there is a possibility of having an inner cusp and an outer cusp specified by the red curves. In addition, by increasing charge, the closed equipotential surfaces also become larger, as predicted by Figure \ref{fig:figsolre}. 

In columns $2$ and $3$ of Figure \ref{fig:figsolre1rot}, we also consider the rotation parameter $a$ for comparison. In general, the deepest analysis reveals the effect of the rotation parameter $a$ on having solutions is not strong compared to $\alpha$ as we see in Figure \ref{fig:figsolre}, but stronger than the effect of the charge parameter $e$. In fact, parameter $a$, like $\alpha$, has a negative effect on having solutions, and for relatively higher acceleration and rotation parameters we do not find any disc structure, unless we add a relatively high charge as far as it is possible.

In Figure \ref{fig:figsolre} we see the distance between the center and the cusp point is an monotonic increasing function of $e$, while it is a monotonic decreasing function of $a$. Therefore, the larger disc structure for bigger charge values $e$ and smaller rotation parameters $a$ is predicted. In other words, on the contrary to $e$, by increasing $a$ and $\alpha$, the centre and the cusp's locations approach one another, and gradually we lose solutions.

Furthermore, in the 2th and 3th columns of Figure \ref{fig:figsolre1rot} by considering rotation, the possibility of having the inner cusp is strongly reduced, while increasing the possibility of the outer cusp, in fact, leads matter to flow outwards. 

In addition, by increasing $\alpha$ and $e$, we obtain closed equipotential surfaces more oriented away from the horizontal axis. Of course the effect of $\alpha$ is more decisive than $e$ in this behavior; namely, as $\alpha$ even slightly increases, the disc deviates from the horizontal axis noticeably, which can be seen more clearly in the following Figures \ref{fig:figsolre1} and \ref{fig:figsolre2}.

In general, as $e$ increases, we expect the matter is concentrated closer to the inner edge of the disc since the slope to reach the cusp is steeper. On the contrary, the higher values of $\alpha$ and $a$  more spread the matter through the disc because the value of the equipotential surface at the centre and at the cusp becomes closer as $a$ increases.

%%%%%%%%%%%%%%%%%%%%%%%%%%%%%%%%%%%% PART 2 RESULTS

In Figures \ref{fig:figsolre1}, we examine the effect of the magnetization parameter $\beta_c$ and the dependency of the disc structure and its orientation on the parameter $e$ in the vicinity of the compact object for a fixed value of acceleration parameter $\alpha$ and vanishing rotation. In the first column we chose high magnetized model, and in the second row relatively high charged one. In fact, comparing columns shows that the magnetization parameter does not influence the disc's geometry; however, it changes the distribution of matter inside the disc and shifts the location of the rest-mass density maximum which is pointed out as the dashed lines in Figure \ref{fig:figsolre1}. In addition, comparing rows show that we have a larger oriented disc for larger values of $e$. Moreover, the matter is more concentrated in the inner part of the disc, as was predicted in the previous Figures \ref{fig:figsolre} and \ref{fig:figsolre1rot}.

In Figures \ref{fig:figsolre2} and \ref{fig:figsolre3} we focus more on the impact of only one parameter $\alpha$ and $a$ on the disc structure, respectively. Figure \ref{fig:figsolre2} presents the profound impact of $\alpha$ on the geometry and orientation of the disc for a fixed value of $e$ and the magnetization parameter for the vanishing rotation. In fact, according to the last row of Figure \ref{fig:figsolre}, the possibility of having solutions for relatively larger values of $\alpha$ depends on having large values for $e$, so the effect of higher $\alpha$ on the disc could be neutralized only partially with the higher charge values.

Figure \ref{fig:figsolre3} shows the dependency of the disc structure on the parameter $a$ for the fixed parameters $\alpha$, $e$, and $\beta_c$. As expected, increasing $a$ changes the disc size and the distribution of matter inside the disc. Furthermore, we do not have an inner cusp for any value of rotation parameters. In addition, increasing $\alpha$ and $a$ shifts the disc farther away from the compact object, contrary to an increase in $e$.

\begin{figure*}
  \centering
\begin{tabular}{cc}
\includegraphics[width=7cm]{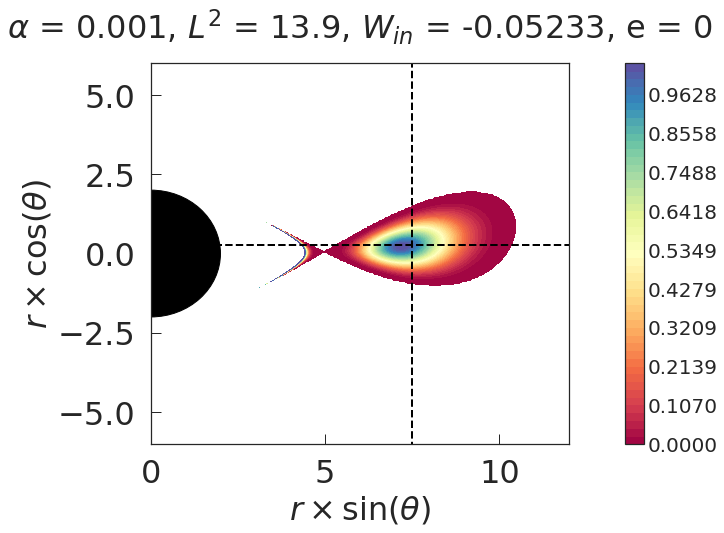}&
\includegraphics[width=7cm]{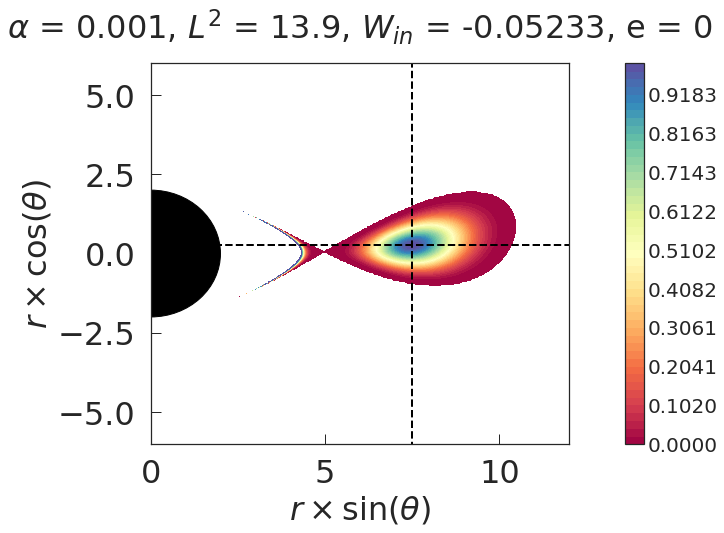}\\
\includegraphics[width=7cm]{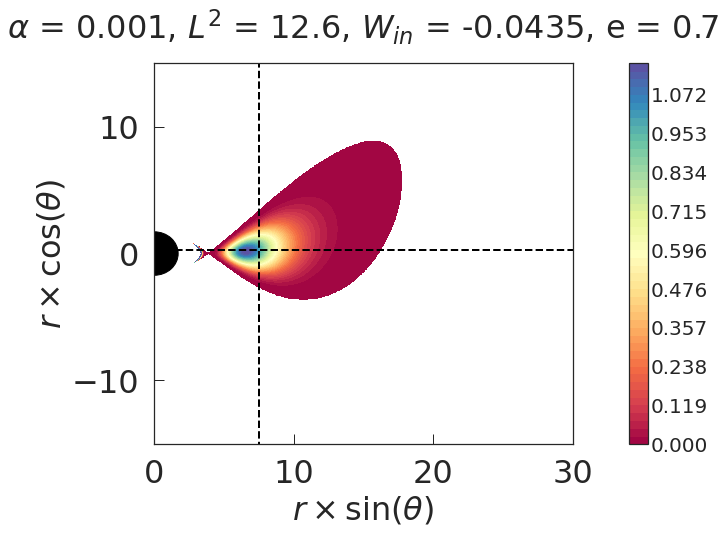}&
\includegraphics[width=7cm]{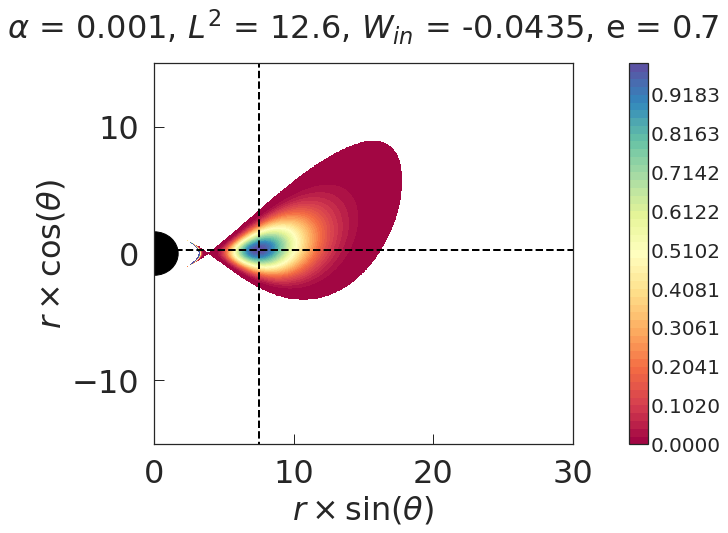}
\end{tabular}
    \caption{Contour map of the rest-mass density of magnetised disc. The dashed lines point the center of the disc located at $r_c=7.5$. The column $1$ shows highly magnetised disc and the column $2$ depicts low magnetised one.}
    \label{fig:figsolre1} % I can do without the label too
\end{figure*}

\begin{figure*}
 \centering
\begin{tabular}{cc}
\includegraphics[width=7cm]{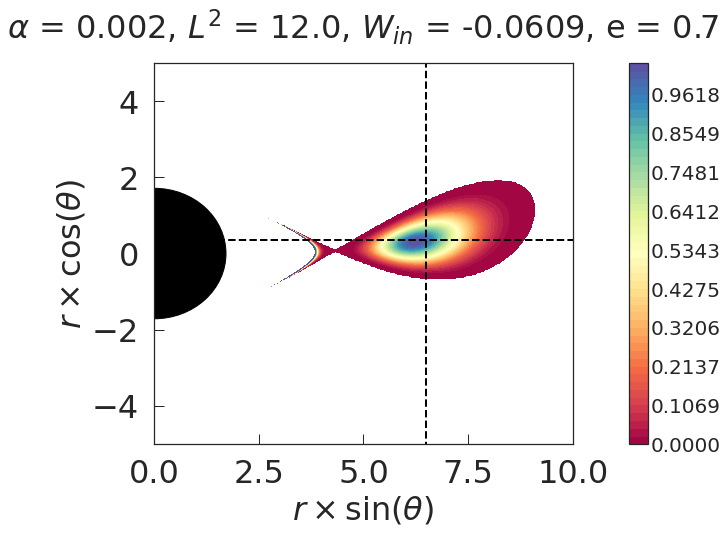}&
\includegraphics[width=7.5cm]{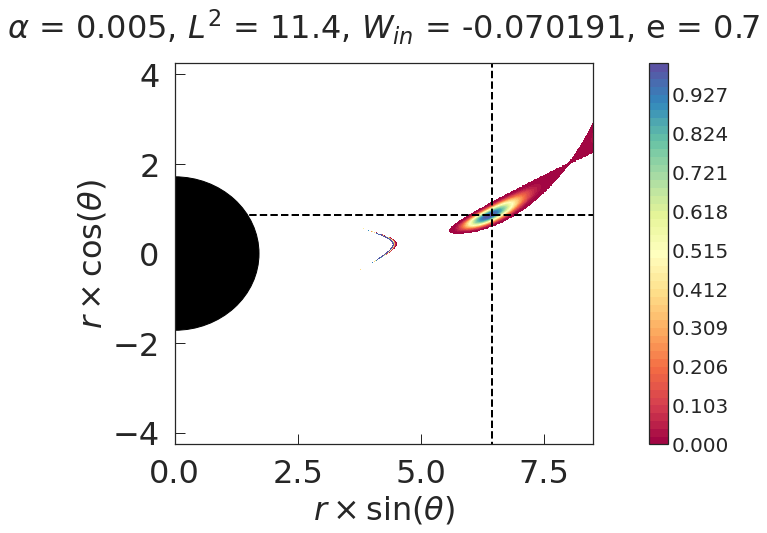}
 \end{tabular}
    \caption{Contour map of the rest-mass density of magnetised highly disc. The dashed lines point the center of the disc located at $r_c=6.5$.}
         \label{fig:figsolre2} 
\end{figure*}

\begin{figure*}
 \centering
\begin{tabular}{cccc}
\includegraphics[width=8.5cm]{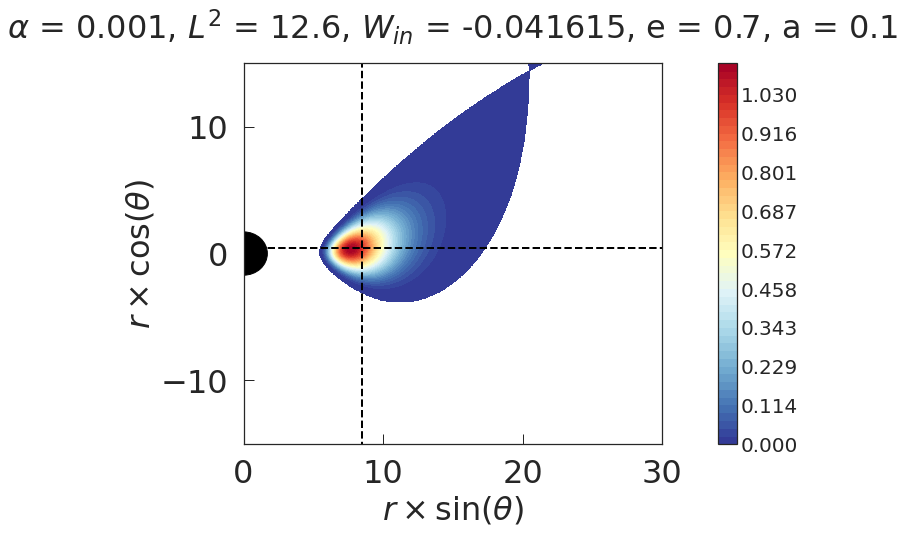}&
\includegraphics[width=8.5cm]{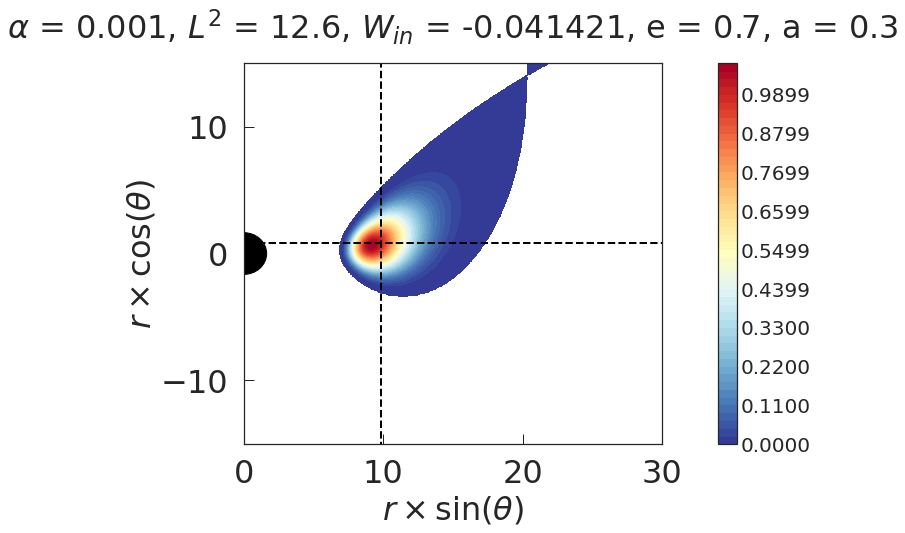}\\
\includegraphics[width=8.5cm]{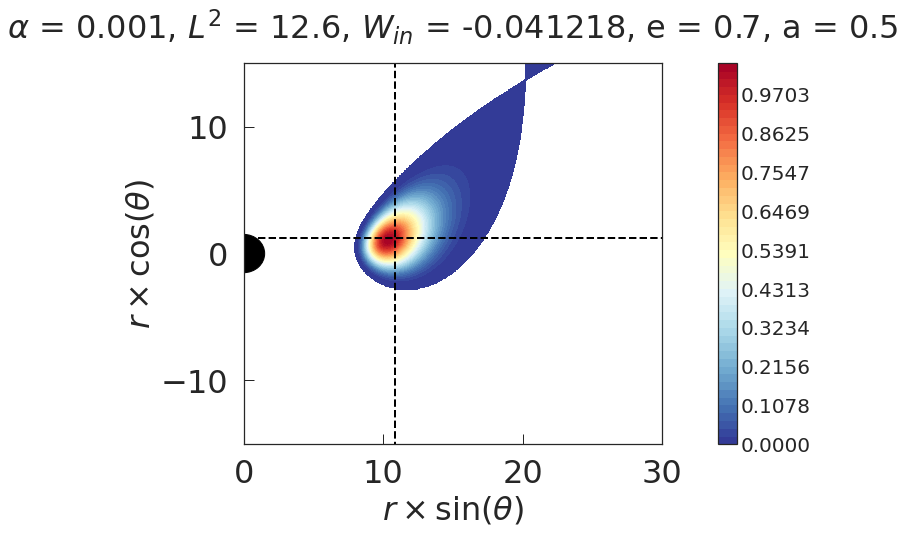}&
\includegraphics[width=8.5cm]{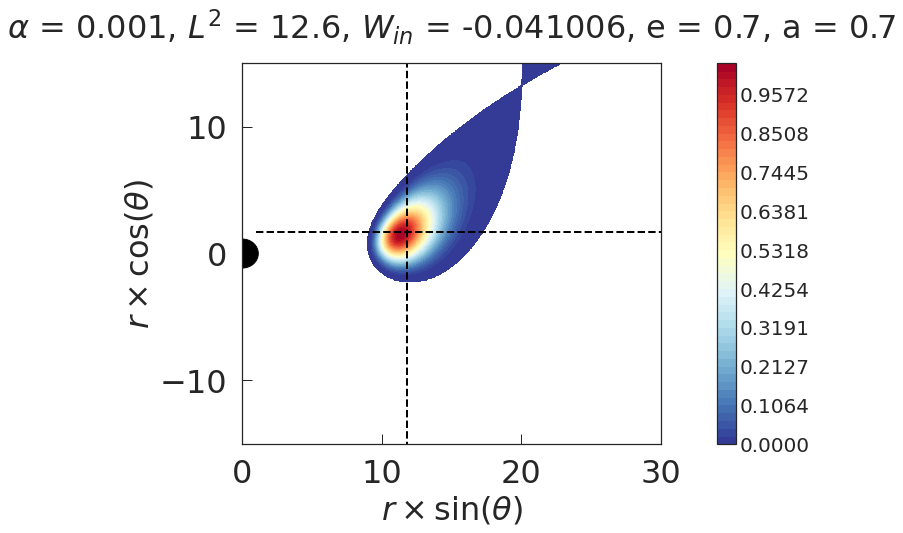}
\end{tabular}
    \caption{Contour map of the rest-mass density of highly magnetised disc for various spin values. The dashed lines point the center of the disc. Those solution have the same parameters ($\alpha, L$ and $e$) of the non-rotating solution given at the bottom left of the Figure \ref{fig:figsolre1}.}
   
      \label{fig:figsolre3} % I can do without the label too
\end{figure*}
As the final point, it is worth mentioning that because of the asymmetry with respect to the equatorial plane, the accretion discs in this space-time, in general, are likely to be unstable even to axis-symmetric instabilities, which is the subject of the following work.
%%%%%%%%%%%%%%%%%%%%%%%%%%%%%%%%%%%%%%%%%%%%%%%%%%%%% SEC 5

\section{Summary and conclusion}\label{sec5}

In this paper, we analysed equilibrium sequences of magnetised, non-self-gravitating discs around a spinning charged accelerating black hole. This solution is described via the generalized $\rm C$-metric family which is briefly explained in the Section \ref{sec2}. In this procedure we considered the approach of Komissarov \citep{Komissarov_2006} to attach a dynamically toroidal magnetic field to the model.

More precisely, we have analyzed the influence of the magnetization parameter $\betac$, charge $e$, rotation $a$, and in particular accelerating parameter $\alpha$ on the structure of the magnetized thick disc model. In one hand, we have shown that changing the magnetization parameter $\betac$ has a noticeable effect on the location and amplitude of the rest-mass density maximum, also distributing the matter inside the disc. The effect of the magnetization parameter is in complete agreement with previous studies using this model e.g \cite{Komissarov_2006}. Furthermore, in this case, the range of isodensity contours increases, which is compatible with the increase of rest-mass density in the inner part of the disc. Indeed, this result remains valid for any chosen value of other parameters.

On the other hand, we have seen the effect of varying the metric parameters: acceleration $\alpha$, charge $e$, and rotation $a$ on the disc's geometry, orientation and its overall shape. We have also shown that the acceleration parameter play a crucial role in the existence and behavior of the solutions in this setup. In conclusion, we can have the thick disc solution only for relatively small values of $\alpha$, and by increasing $\alpha$, the disc structure becomes smaller and slowly oriented away from the horizontal axis and gradually vanishes. Furthermore, higher values of $\alpha$ shift the disc farther away from the blackhole. Additionally, $a$ has a similar effect but weaker on the structure; by increasing $a$, the disc becomes thinner and smaller and more oriented concerning the horizontal axis until it vanishes completely. On the contrary to these two parameters, an increase in $e$ increases the disc size and possibility of having solutions.

In addition, we have seen that $e$ changes the distribution of matter inside the disc in the opposite way of $\alpha$ and $a$. Besides, increasing $\alpha$ and $a$ shifts the disc farther from the compact object, contrary to an increase in $e$. However, we should emphasize that the strength of the parameters are not the same; among these three parameters, $\alpha$ has a more substantial and $e$ has the weaker effect on the disc structure, in comparison. In general, the impact of the charge parameter is the inverse of $\alpha$ and $a$ in any aspects regarding the disc properties.

As a further step of this work, the time-like circular motion can be studied. The instability of the resulting solution also deserves a proper analysis. It is also of some interest to apply these models as the initial conditions in the numerical simulations and test their ability to account for observable constraints of astrophysical systems.

%%%%%%%%%%%%%%%%%%%%%%%
\section* {Acknowledgements}
S.F. and A.T thanks Torben Frost for a useful discussion. S.F. acknowledges the Cluster of Excellence EXC-2123 Quantum Frontiers -- 390837967 by the German Research Foundation (DFG). A.T. thanks the research training group GRK 1620 ”Models of Gravity”, funded by the German Research Foundation (DFG). V.K. acknowledges the Czech Science Foundation (21-11268S).

\bibliographystyle{unsrt}
\bibliography{ThickCmetric}
\end{document}